%% file: main.tex
\documentclass[10pt, journal]{IEEEtran}

\newcommand{\ignore}[1]{}
\usepackage[boxed]{algorithm}
\usepackage{fancyhdr}
\usepackage[normalem]{ulem}
\usepackage[hyphens]{url}
\usepackage{microtype}
\usepackage{graphicx}
\usepackage{xcolor}
\usepackage{xspace}
\usepackage{multirow}
\usepackage{authblk}
\usepackage{graphicx}
\usepackage{array}
\usepackage{booktabs}
\usepackage{epstopdf}
\newcommand{\pecim}{{Eva-CiM}\xspace}

\newcommand{\eat}[1]{}

\ifCLASSINFOpdf
\else
\fi

\begin{document}
%
\title{Eva-CiM: A System-Level Performance and Energy \\ Evaluation Framework for \\ Computing-in-Memory Architectures}
%
%
%

\author{Di~Gao,~\IEEEmembership{}
        Dayane~Reis,~\IEEEmembership{}
        Xiaobo~Sharon~Hu,~\IEEEmembership{Fellow,~IEEE}
        Cheng~Zhuo,~\IEEEmembership{Senior~Member,~IEEE}
        \thanks{Manuscript received May 27, 2019; revised September 30, 2019. This paper was recommended by Associate Editor A. Coskun.
        This work was supported in part by the NSFC under grant 61974133 and 61601406, Zhejiang Provincial Key R\&D program under grant 2020C01052, National Key R\&D Program of China under grant 2018YFE0126300, and in part by the NRI Program of SRC, by the NSF under grant 1640081, and EXCEL, an SRC-NRI Nanoelectronics Research Initiative under Research Task ID 2698.004, and Asian Research Grant from the University of Notre Dame.
         \textit{(Corresponding author: Cheng Zhuo.)}}
        \thanks{D. Gao and C. Zhuo are with the department of Information Science \& Electronic Engineering, Zhejiang University, Hangzhou 310027, China (e-mail: czhuo@zju.edu.cn).}
        \thanks{D. Reis and X. S. Hu are with the University of Notre Dame, Notre Dame, IN 46556, USA.}

        }

\maketitle

\begin{abstract}
Computing-in-Memory (CiM) architectures aim to reduce costly data transfers by performing arithmetic and logic operations in memory and hence relieve the pressure due to the {\em memory wall\/}. However, determining whether a given workload can really benefit from CiM, which memory hierarchy and what device technology should be adopted by a CiM architecture requires in-depth study that is not only time consuming but also demands significant expertise in architectures and compilers. This paper presents an energy evaluation framework, \pecim, for systems based on CiM architectures. \pecim encompasses a multi-level (from device to architecture) comprehensive tool chain that leverages existing modeling and simulation tools such as GEM5~\cite{gem5}, McPAT~\cite{mcpat} and DESTINY~\cite{destiny}. To support high-confidence prediction, rapid design space exploration and ease of use, \pecim introduces several novel modeling/analysis approaches including models for capturing memory access and dependency-aware ISA traces, and for quantifying interactions between the host CPU and the CiM module. \pecim can readily produce energy and performance estimates of the entire system for a given program, a processor architecture, and the CiM array and technology specifications. \pecim is validated by comparing with DESTINY~\cite{destiny} and \cite{jain18_sttcim}. \pecim enables analyses including the system-level impact of CiM-supported accesses, whether a program is CiM-favorable as well as the pros and cons of increased memory size for CiM. \pecim also facilitates exploration of different design configurations and technologies.
\end{abstract}


%
\IEEEpeerreviewmaketitle

\input{introduction.tex}

\input{background.tex}

\input{overview.tex}

\input{analysis.tex}

\input{modeling.tex}
\input{profiling.tex}

\input{exploration.tex}

\section{Conclusions}
This paper presents a system-level evaluation framework, \pecim, to predict performance and energy consumption of CiM based systems for different architectures/configurations, technologies and benchmarks. Unlike prior work, \pecim relies on a novel IDG based analyzer to automatically detect offloading candidates for CiM and uses multi-level modeling to provide comprehensive evaluations of a CiM based system. \pecim is capable to conduct quantitative investigations and rapid design exploration, thereby further establishing the feasibility of CiM for wide adoption in the near future.

We validate \pecim with two existing works \cite{jain18_sttcim} and \cite{destiny} with respect to both access count and energy consumption. We then investigate various data sensitive benchmarks to explore the count of CiM-supported memory accesses, system speedup and its energy improvement over a non-CiM based design. It is found that for a system with a multi-level memory hierarchy, data sensitive benchmark is not necessarily CiM-sensitive. Moreover, it is not necessarily beneficial to CiM with larger memory sizes due to the increased energy per CiM operation by the CiM module itself. Finally, \pecim evaluates the impact of different architecture configurations and technologies, and the results show that SRAM CiM can provide 1.0-1.5$\times$system speedup, 1.3-6.0$\times$ energy improvement for SRAM and 2.0-7.9$\times$ for FeFET-RAM, respectively.


%




\ifCLASSOPTIONcaptionsoff
  \newpage
\fi



%

\bibliographystyle{IEEEtran}
\input{main.bbl}

%




\end{document}

%% file: introduction.tex
\section{Introduction}
\label{sec_introduction}

\IEEEPARstart{W}{}ith the rapid growth of Internet of Things (IoT), the era of "Big Data" is upon us and features massive data transfers between processor and memory \cite{paul2014}. The efficiency of the conventional Von Neumann architecture is severely restricted by its limited bandwidth and increasingly complex interconnects, which result in significant energy and latency overhead for data movement. For instance, the energy spent on transferring 256 bits from main memory to the processor is estimated to be 200$\times$ higher than the energy for one floating point operation \cite{keckler11}.

Researchers have long been aware of the inefficiency of conventional architectures for data movements~\cite{Gokhale95_tc, patterson97_isscc}, and spent significant efforts on trying to bring computation closer to  (or even inside) memory ({\em e.g.\/}, \cite{oskin98, mai00, hmc13}. Such architectural designs that places logic and memory close to each other are often referred to as near-memory computing (NMC), processing in memory (PIM), or computing in memory (CiM). NMC reduces the energy and latency associated with memory accesses by placing processing units {\em close to\/} the memory. PIM or CiM, often found to have interchangeable meanings in the literature, are architectures that integrate certain logic and arithmetic operations directly in either the memory cells or memory peripherals, in order to lower the number of memory references made by the processor. For the sake of conciseness, from this point on, we simply use CiM to refer to the PIM/CiM architectures defined above.



Recent works ($e.g.$, \cite{jain18_sttcim, farmahini15, zhang14, tesseract_isca_15, Akin_isca_2015, Ahn_ISCA_2015, hshieh16, prime_isca_16, aga17, agrawal18,  reis18_fefetcim, Liu_MICRO_2018}) in both CMOS SRAM and emerging non-volatile memories (NVMs) have demonstrated various CiM designs at different levels of memory hierarchy. These designs allow computation to occur exactly where data resides, thereby reducing energy and performance overheads associated with data movement. For example, cache based CiM in \cite{aga17} achieves 2.4-9$\times$ energy saving on text processing scenarios. Meanwhile, NVM based designs such as \cite{jain18_sttcim, prime_isca_16, reis18_fefetcim} can improve energy saving up to two orders of magnitude when functioning as a co-processor on neural network benchmarks.  

While CiM is found to be a powerful and promising alternative with various design options, such variety also complicates the design process. Designers are confronted with several important questions when designing CiM:
\begin{itemize}
    \item How much can an application program benefit from a CiM based {\em system}?
    \item At which level of memory hierarchy should one place the CiM?
    \item Which technology should be used for CiM?
\end{itemize}
%
There are some prior efforts attempting to address the above questions. However, they suffer limitations in several aspects.
   
   {\bf Overall system evaluation}: Most CiM works ($e.g.$,~\cite{reis18_fefetcim, chowdhury18_impspintronics, xie18_aimaes}) focus on the CiM module without considering the host CPU or use an emulation platform consisting of a simple host CPU \cite{jain18_sttcim}. Interactions between the host and the CiM as well as the complete memory system can be rather complex and the impact on the energy/performance of the overall system can be quite significant \cite{zhuo2018layout}.
    
   {\bf Offloading candidate identification}: Offloading candidate here refers to a code snippet, a function or an instruction that can be offloaded from the host CPU to the CiM module for execution. Most prior solutions do not provide instruction set architecture (ISA) nor compiler support to automatically determine offloading candidates. Designers have to either manually identify the code snippets for CiM from the entire benchmark ($e.g.$, ~\cite{Ahn_ISCA_2015, prime_isca_16, aga17}), or select specific instruction groups for offloading to the CiM unit for execution ($e.g.$, ~\cite{jain18_sttcim, reis18_fefetcim}). In the latter two works, memory accesses triggered by a CiM module with custom instructions are identified at compiling time. Two operands fetched from the same level of memory are replaced by one CiM instruction. 
    The method cannot be generalized to systems with multi-level caches as it assumes ideal locality and dependence.
    
    {\bf Multi-level modeling}: CiMs based on different devices, circuits and micro-architectures have been proposed. However, there is no uniform framework to compare design options at different levels. Though some existing work such as~\cite{reis18_fefetcim} has compared CiMs implemented with different technologies, the comparisons are hand crafted and cannot be easily adapted to different memory hierarchies. 
 In this paper, we present an architectural evaluation framework, \pecim, that overcomes the above limitations and is able to reliably predict energy consumption and performance of any systems containing a CiM module. The major contributions of this work are summarized below.
     
 We propose a novel trace-driven analysis method to extract data dependencies and identify offloading candidates. The method is built on an instruction dependency graph model augmented with memory access information. The analyzer is integrated into GEM5 \cite{gem5} and hence can readily work with different architectures, compilers and development options. 
    
We leverage a comprehensive tool chain from device to architecture to build a multi-level CiM model. We employ GEM5 as the backbone of the framework to fully capture the effects of both the host CPU and the complete memory hierarchy. We further design and embed a probe-based simulation inside GEM5 to collect the necessary information for offloading candidate selection at the application layer. We extend McPAT~\cite{mcpat} by including a CiM module obtained from SPICE~\cite{spice} and DESTINY~\cite{destiny} to provide the architectural-level energy profiling capability. \pecim is validated by comparing with DESTINY~\cite{destiny} and \cite{jain18_sttcim}.

 We employ \pecim to investigate the three questions raised earlier. Unlike prior works which typically assume ideal data locality and regular memory accesses, \pecim is able to find operations that are offloadable to CiM under realistic architecture and compiler settings and thus avoid being overly optimistic. Furthermore, we use \pecim to quantitatively estimate the system speedup, evaluate the energy saving of CiM not only due to the reduced memory accesses but also the lower computational loads on the host. Last but not least, we conduct design space explorations on different technologies, system configurations to illustrate the design options that maximize the CiM benefits for a set of benchmarks. 
     
 \pecim enables us to make the following findings that are either different from or have not been seen in the conclusions presented in prior works: (i) In a general purpose CiM based system with complete memory hierarchy, the number of possible CiM-supported accesses is similar as the one for regular access; (ii) Data intensive benchmarks are not necessarily always CiM sensitive. The sensitiveness depends on both benchmark characteristics {\em and\/} system architecture; (iii) Energy wise, larger memory size is not necessarily helpful for CiM due to the increased energy per CiM operation.

%% file: background.tex
\section{ Background and Related Works}
\label{sec_background}

Below, we review the backgrounds on CiM and existing works for CiM modeling and estimation.

\subsection{Computing in Memory}
To address the performance gap between processing and memory access, there have been significant efforts aiming at bringing computation closer to memory. Earlier works, $e.g.$,~\cite{oskin98, mai00}, focused on devising architectures that combine processing cores with dynamic random-access memory (DRAM) modules. These architectures generally belong to the category of 
NMC \cite{singh-18dsd}. However, practical concerns regarding the successful integration of DRAM and processing units into the same chip have hindered the advancement of such NMC systems for many years. Recently, the advent of 3D stacking memories using massive through-silicon-vias (TSVs) provides larger bandwidth while allowing the integration of logic and memory into a stacked chip ($e.g.$,~\cite{hmc13, farmahini15, zhang14, hshieh16, jedec13}). 

To bring processing and memory even closer, 
CiM --- where processing is done in the memory array --- is gaining a lot of attention recently in both academia and industry. This growing interests largely attribute to the needs of data-intensive IoT applications, and the advances in circuits and device technologies \cite{zhuo2019noise,luo2018noise}. Many design alternatives exist for CiM, which vary in circuit style, supported operations, device technologies, location in the memory hierarchy, application targets, $etc.$ The most extreme design of CiM is to embed logic operations within each memory cell~\cite{kvatinsky14_magic, zabihi_spintronicCRAM}. Though such designs eliminate the memory access overhead, their negative impact on memory density prevents them to be widely employed. 
Another CiM design style is modifying the peripheral circuitry of the memory array (either SRAM or DRAM) to realize logic and arithmetic operations. This design style offers a good balance between memory density and processing efficiency. 
For example, some works proposed to modify the peripheral circuitry, $e.g.$, the sense amplifiers (SAs) of caches, to enable CiM~\cite{aga17,agrawal18}, while others accomplish CiM through supporting bulk bit-wise operations using the features of DRAM~\cite{Seshadri_CoRR_2016}. 

Progress in emerging non-volatile device is further fueling the development of CiM. Specifically, non-volatile resistive RAMs (ReRAMs), phase changing memory (PCM), spin-transfer-torque magnetic RAMs (STT-MRAMs), and ferroelectric field effect transistor-based RAMs (FeFET-RAMs) offer high density, good scalability, and low power, making them natural candidates for realizing CiM memory architectures.
For instance, there have been a number of recent efforts investigating CiM-capable NVRAMs --- employed as either cache or main memory --- for various applications. References~\cite{jain18_sttcim,reis18_fefetcim, li16_pinatubo} study the use of NVM with a re-designed SA to perform a subset of logic and arithmetic operations.  In~\cite{imani17_rramtcam, xyin17_fefettcam}, NVMs are used in content addressable memory (CAM) to support parallel search while reducing data transmission in data-intensive IoT applications. Many recent works also employ NVM-based circuits for neural network acceleration by directly executing matrix-vector multiplication \cite{liu2018multi} within the memory array \cite{prime_isca_16,prezioso15, jerry17_dnn}.
Reference \cite{imani19_floatpim} further improves energy efficiency for neural network training and testing through the implementation of a fully-digital scalable CiM architecture.

\begin{figure*}[htb]
    \centering
    \includegraphics[width=18cm]{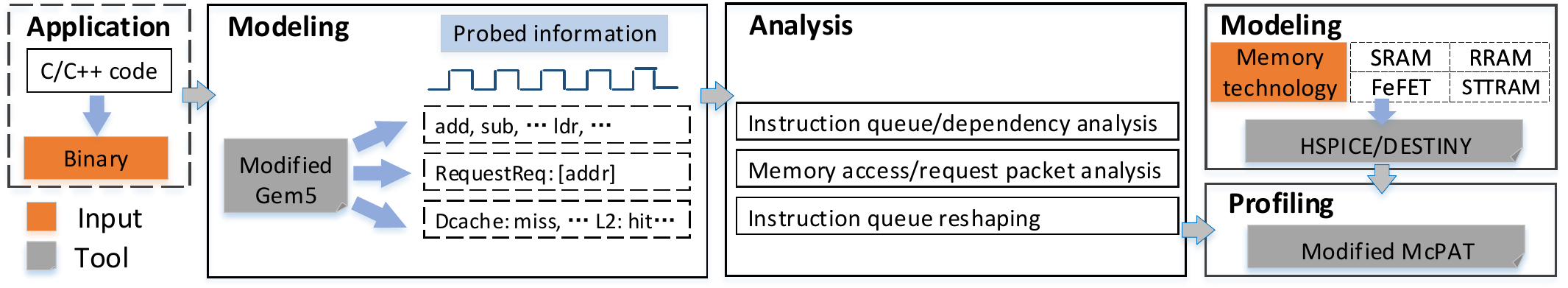}
    \caption{Overview of the \pecim framework: data flow, tool chain and architecture.}
    \label{fig_overview}
\end{figure*}

This paper focuses on systems that contain a host CPU and a CiM module that can be placed in any level of cache hierarchy. Furthermore, the CiM module can be implemented in different technologies/styles and support different instruction sets. 



\subsection{Related Work}
\eat{
Given the large CiM design space encompassing device, circuit, micro-architecture and ISA  variations, 
it is a daunting task for designers to predict how the choice of CiM design options affects the overall system (including both the host and CiM unit) energy and performance for a given application. 
}

Many system-level simulators, such as GEM5~\cite{gem5}, zsim~\cite{zsim}, Sniper~\cite{sniper}, $etc.$, only cover architectural details for general purpose processor simulation. On the other hand, some existing CiM efforts have attempted to compare different CiM design options by evaluating the energy/performance of the CiM module or accelerator alone \cite{jain18_sttcim, zhang14, reis18_fefetcim, chowdhury18_impspintronics, xie18_aimaes}. In all cases, the focus is on estimating energy savings due to (i) a lower number of memory accesses in CiM-enabled systems, and (ii) the inherently high internal bandwidth of the memory architecture. Though these comparisons are important in understanding the pros and cons of different CiM designs, they cannot predict the overall benefit offered by CiM based systems since they do not consider how many instructions can actually be offloaded to the CiM module and the effect of such offloading has on the host CPU.

Recent works~\cite{jain18_sttcim, Ahn_ISCA_2015} have tried to estimate the benefits of CiM to data-intensive applications by using custom CiM instructions. The work in \cite{jain18_sttcim} extends the original ISA of the Intel Nios II processor \cite{intel_niosii} with custom CiM instructions. The memory module is assumed to be a small SPM. At compiling time, memory accesses of given application benchmarks are categorized into (i) writes ($W\!R$), (ii) non-convertible ($N\!C$) reads, and (iii) CiM convertible ($CC$) reads, $i.e.$, reads triggered by CiM instructions. 
The evaluation assumes that every two $CC$ reads could be effectively replaced by one CiM instruction. 
Although the approach provides a good insight on the benefits of CiM to systems with a single-level non-cacheable memory, issues like memory hierarchy and locality of data are not taken into consideration. Furthermore, the impact of CiM instructions on the host processor is not studied. Therefore, the method may not be generalized to estimate the overall system-level CiM benefit for most real-world systems that leverage multi-level caches. 

As an alternative, the work in \cite{Ahn_ISCA_2015} implements a set of custom instructions in a x86-64 architecture. Different from \cite{jain18_sttcim}, multi-level caches are considered in the evaluation. Furthermore, the work proposes taking the data locality into consideration in order to determine whether it is worthwhile to offload potential CiM instructions to the memory unit. Note that the in-memory operations are called by atomic instructions specific for HMC model integrated to the system. However, instead of looking into the memory access breakdown to define the instructions offloaded to the CiM module, the method assumes that the system designer has enough knowledge about the application, and can manually insert CiM-enabled macros into the appropriated code snippets. An obvious limitation to the approach is that it does not offer a systematic way to locate all the possible places in which CiM-enabled macros could be inserted, which inevitably underestimate the benefits of CiM. Different from most current works, \cite{aga17} explores CiM in three levels of SRAM cache hierarchy and completes the control flow inside cache in the absence of data locality. However, its limitation is the same as \cite{Ahn_ISCA_2015}, which requires customized benchmarks for data locality. 

Our work aims to address the needs for a framework to help designers predict how the choice of CiM design options affects the overall system (including both the host and CiM module) energy and performance for a given application. We leverage several existing memory and micro-architecture modeling tools. (Note that these modeling tools focus on either one particular layer of memory or general microprocessors, thus cannot be easily extended for system-level evaluation of CiM based systems.) Specifically, we use DESTINY \cite{destiny} to estimate energy at array-level for L1/L2 levels of cache, modifying it to support the particularities of specific CiM designs, $i.e.$, customized sense amplifiers \cite{aga17, reis18_fefetcim} and memory cells \cite{ reis18_fefetcim}. DESTINY is an open-source tool to simulate 2D and 3D memory arrays, which utilizes the 2D modeling framework of NVSim~\cite{dong12_nvsim} for SRAM and NVMs, and the 3D framework of CACTI-3DD~\cite{chen12_cacti3dd}. Besides, McPAT~\cite{mcpat}, an integrated power, area, and timing modeling tool, is modified and used to evaluate different components ($e.g.$, CiM, core, caches) at architecture-level. 



%% file: overview.tex
\section{Overview of Eva-CiM}\label{sub1}

Our proposed \pecim framework adopts a combined simulation and analysis approach to accomplish performance, energy estimation for an entire CiM based system. Besides leveraging several existing architecture and circuit simulators, \pecim builds its own models at different design levels. 
Fig. \ref{fig_overview} depicts the overall flow, structure and tool chain of \pecim, which consists of three stages: modeling, analysis and profiling. \pecim takes as input (represented by the orange boxes) the binary of a given application, the device and CiM array parameters for the CiM module, and output the overall evaluation results. The simulation and analysis tools used in \pecim are shown as grey boxes. The current version of \pecim supports two technologies, including SRAM and FeFETs, meanwhile more technologies can be readily added later.
\begin{figure}[htb]
    \centering 
    \includegraphics[width=8.5cm]{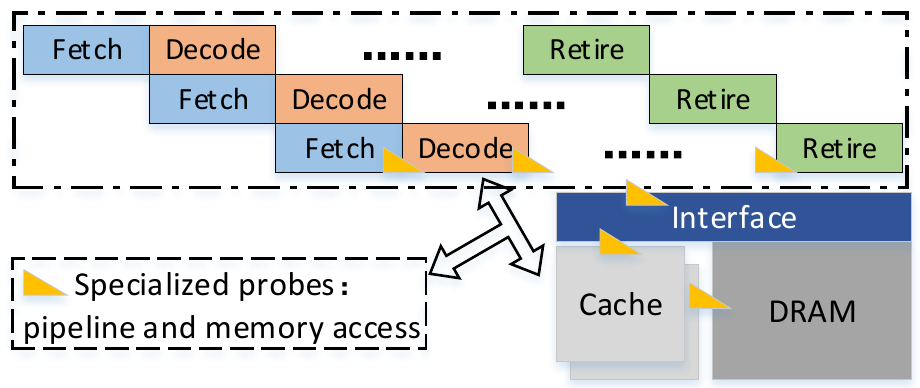}
    \caption{GEM5 simulation and specialized probes to extract information.}
    \label{fig_simulation}
\end{figure}

The \textbf{modeling} stage in \pecim aims to construct models to be used in the analysis stage. Specifically, \pecim uses two models: the application model and device/CiM array model. The application model captures when and where instructions are executed and memory accesses occur. This information will be used in the analysis stage to determine offloading candidates. At the application level, \pecim can take any binary that is compiled from GEM5-compatible general-purpose or customize compilers~\cite{gem5}. The benchmark binaries are fed to a modified GEM5 and goes through fetching, decoding and commit with specialized probes to extract the pipeline and memory access information, as illustrated in Fig. \ref{fig_simulation} (details in Section \ref{subsec_app}). 

The device/array model describes energy consumption by individual
CiM operations, such as CiM-OR, $etc.$ (details in Section \ref{subsec_device}). This can be achieved by SPICE simulation~\cite{spice} if the netlist is available, or users can break down an atomic CiM operation into its micro-operations, $e.g.,$ cell access, SAs, $etc.$, and then use DESTINY~\cite{destiny} with pre-calibrated energy data to compute the energy for the atomic operation. Unlike application level simulations, the device/array level simulation is conducted per technology to extract the models.

The \textbf{analysis} stage investigates data dependence and locality and decides the offloading candidates (details in Section \ref{sec_analysis}) and is the cornerstone of \pecim. Several new ideas are introduced here, $e.g.$, the instruction-dependency-graph used to organize and interpret the instruction execution and memory access information obtained from the modeling stage. The key analysis conducted by \pecim include: (i) Committed instruction queue and dependency analysis to automatically detect the underlying offloading patterns; (ii) Memory access and request packet content analysis to determine the particular cache level for CiM; (iii) Instruction trace reshaping to enable system-level profiling.

The \textbf{profiling} stage estimates performance and energy consumption based on the results from the analysis stage. For profiling, \pecim employs modified McPAT~\cite{mcpat} and use reshaped instruction queue statistics to analytically compute the energy overheads of different components in the system (details in Section \ref{subsubsec_system}).
The tool chain of \pecim shown in Fig. \ref{fig_overview} leverages four existing tools. HSPICE~\cite{spice} and DESTINY~\cite{destiny} are used for memory cell and array modeling. GEM5~\cite{gem5} is modified to include specialized probes to model applications, while McPAT \cite{mcpat} is enhanced to provide CiM system profiling. 
 
\pecim is not limited to a particular technology/architecture nor development environment, $e.g.,$ compiler. In other words, it provides a unified and ease-to-use interface for designers to evaluate the pros and cons of CiM based systems. Hence, \pecim can support design space exploration (details in Section \ref{sec_design}) including (but are not limited to):
\begin{itemize}
    \item Study of the ratio of CiM-supported instructions over regular memory accesses to decide if an application is CiM-friendly or not;
    \item Comparison of various device technologies;
    \item Determination of the best memory hierarchy level for the CiM module given the concerned applications.
\end{itemize}

In the following sections, we will detail the design of \pecim by first discussing the analysis stage as it is the core of \pecim and then the modeling and profiling stage. We finally present interesting findings obtained by applying \pecim to show the capability of \pecim and the necessity of such a framework for CiM based system.

%% file: analysis.tex
\section{Analysis}\label{sec_analysis}

A key task in evaluating the benefit of including a CiM module to the overall system is to determine what can be offloaded to the CiM module. 
As discussed in Section~\ref{sec_background}, most prior works make contributions either by manual analysis or by limiting the system to a specialized simple architecture to cater to condition of offloading candidates. Clearly this requires non-trivial efforts for design space exploration especially when considering complex programming and architecture development environments. In view of prior progress that proves the availability of CiM paradigm from circuit level to higher level, we focus on analyzing suitable offloaded instructions based on real-world applications and then evaluate their effects.
As discussed in Section \ref{sub1}, \pecim presents a unified interface and development environment to hide the aforementioned complexities inside the framework, thereby allowing convenient and efficient design exploration. Specifically, \pecim embeds a trace-driven analyzer in GEM5~\cite{gem5} to analyze the committed instructions, and then identifies the proper candidates as well as data locality for CiM. In other words, programmers can rely on the framework without manually identifying the critical functions in the code for CiM or dealing with complex development environment.
To enable the proposed trace-driven analyzer, we need to answer the following key questions: 
\begin{itemize}
\item What instruction patterns can be offloaded to memory to maximize the benefit of CiM?
\item How to analyze the program dependencies to identify and select the proper patterns?
\item How to reshape the instruction queue after offloading to facilitate system level profiling? 
\end{itemize}

\subsection{Offloading Candidate Selection}

\begin{figure}[b]
    \centering 
    \includegraphics[width=8.45cm]{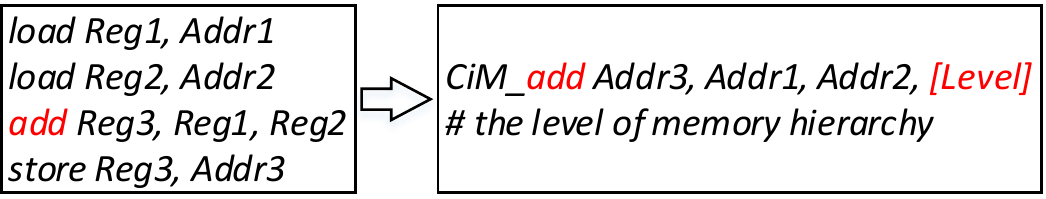}
    \caption{An example of Load-Load-OP-Store pattern.}
    \label{pattern}
\end{figure}

We first examine what instructions patterns are CiM suitable and can be offloaded to the CiM module.
In general, an instruction that is suitable for CiM is featured with source operands fetched from memory and destination operand stored to memory. One common pattern that prior works \cite{jain18_sttcim, reis18_fefetcim} rely on is a sequence of Load-Load-OP-Store instructions, as on the left of Fig. \ref{pattern}, in which two load operations obtain the source operands, one ``OP" instruction (``add" in the figure) conducts a particular operation, and one store operation saves the result. Then this sequence can be replaced by a CiM instruction, $e.g.$, in-cache operation $CiM\_add$ as on the right of Fig. \ref{pattern} \cite{jain18_sttcim}. 

\begin{figure}[tb]
    \centering
    \includegraphics[width=5.8cm]{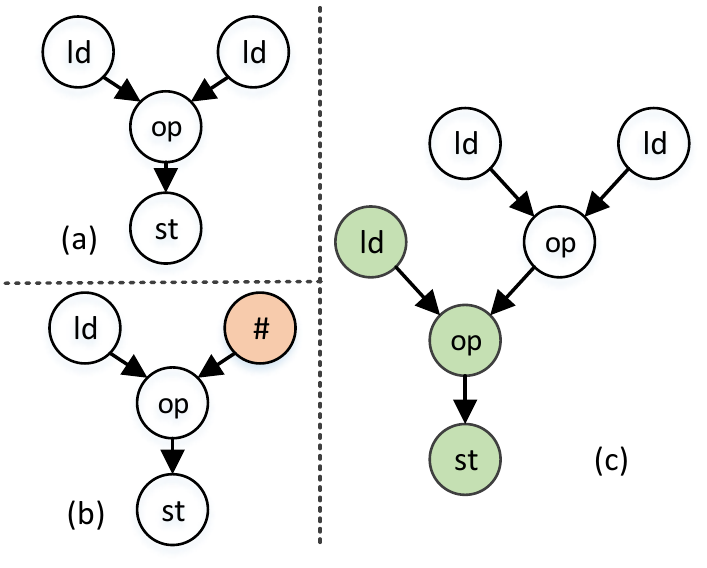}
    \caption{IDGs for the original Load-Load-OP-Store pattern and its variants.}
    \label{newpattern}
\end{figure}

However, 
due to compiler optimization and usage of intermediate resources ($e.g.$, integer and floating registers), such an exact Load-Load-OP-Store pattern rarely occurs during instruction execution. Instead, the Load-Load-OP-Store pattern adapts to multiple variants, as shown in Fig. \ref{newpattern}(b),(c), which are different but all suitable for CiM. Unlike the regular pattern in Fig. \ref{newpattern}(a), Fig. \ref{newpattern}(b) replaces one source operand with an immediate value while Fig. \ref{newpattern}(c) continues using the output before it is stored back to memory. Moreover, it is not uncommon to have a combination of two or more such patterns to form a large CiM-suitable pattern. 

In order to capture the complex dependencies among instructions and help identify CiM-suitable patterns, we resort to a graph model, called Instruction Dependency Graph (IDG). In IDG, a ``node'' is an instruction and a directed ``edge'' indicates the execution order of the two instructions with data dependency. Fig. \ref{newpattern} are examples of three IDGs. 
If one straightforwardly builds an IDG for all the instructions being fetched, the IDG would be overwhelmingly complicated.
In Section~\ref{subsec_idg}, we will present an approach to construct a more manageable IDG for a given program.

\begin{figure}[tb]
    \centering
    \includegraphics[width=8.5cm]{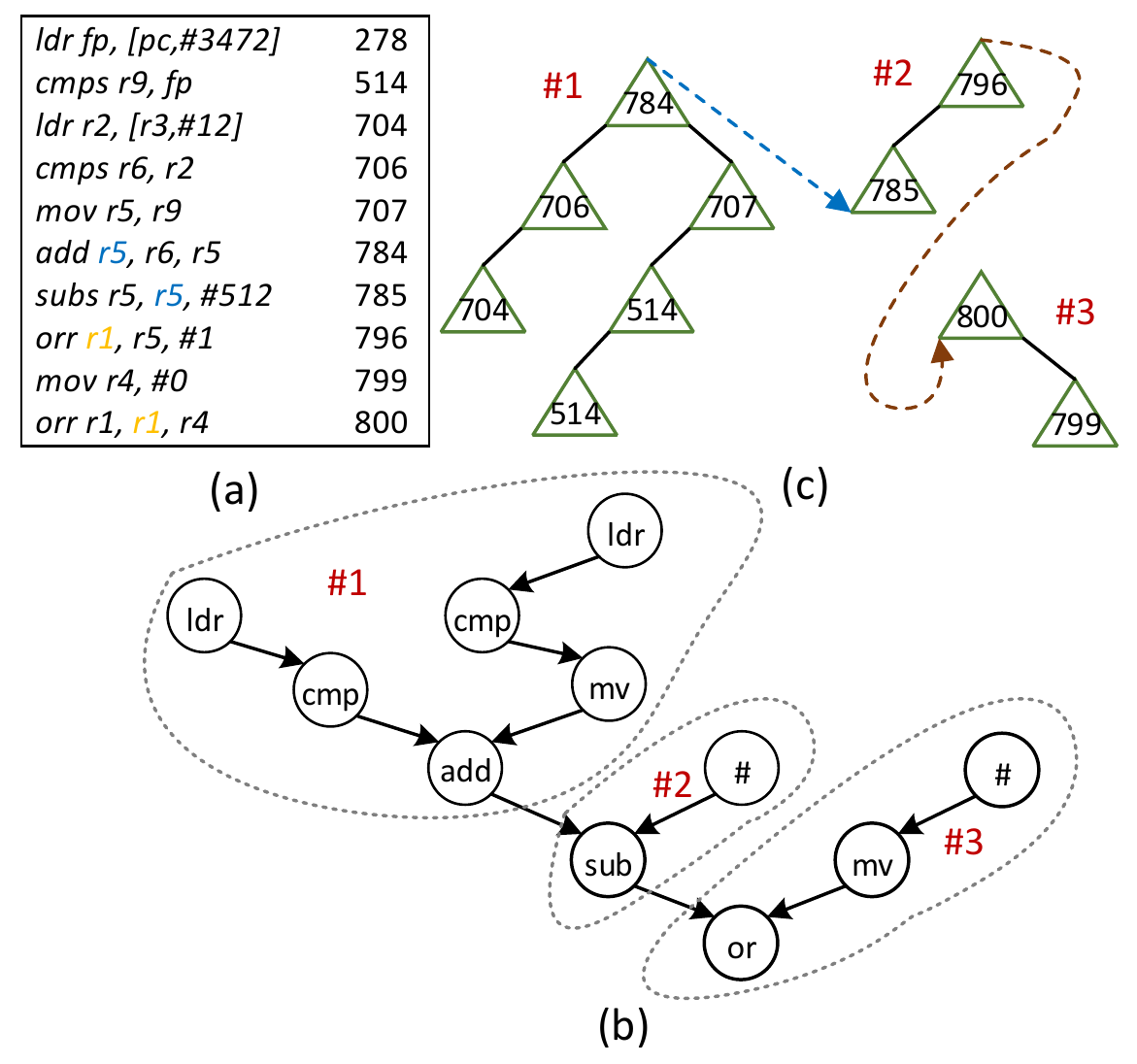}
        \caption{An example of extracting offloading candidates from the committed instruction queue: (a) instruction snippet, (b) the corresponding IDG and its partition, (c) the resulting CiM offloading candidates where each triangle represents one instruction and the number inside the triangle represents its sequence index in the queue.}
    \label{treeexample}
\end{figure}

Besides the instruction execution patterns captured in an IDG, memory access information is also crucial for offloading candidate selection. For example, the operands of a candidate CiM operation should be from the same memory bank. 
Thus for the level of memory of a leaf node instruction, we need to check if its request address is within the access address of memory objects and then obtain the corresponding Miss-Status Handling Register (MSHR) state~\cite{gem5}. We can do such a procedure repeatedly until we find the memory hierarchy level that stores the data. Depending on the operations that the CiM unit supports, one or multiple sub-trees can be identified as offloading candidates from one IDG tree. Fig. \ref{treeexample} presents a simple example of the procedure to select the offloading candidates, where the IDG tree contains three sub-trees that are identified as proper offloading patterns for CiM. 


\begin{algorithm}[b]
\caption{Algorithm for offloading pattern selection.}
\setlength{\parindent}{0pt}
 	$\textbf{Procedure:}$ \  Offloading candidate selection \\
	$\textbf{Input:}$ \ I-state for all instructions \\ 
	$\textbf{Output:}${ \ CiM operations }\\
    1.\ Build register usage table $RUT$ and index hash table $IHT$;\\
    2.\ Build IDG trees for the committed instruction queue $CIQ$; \\
	3.\ Partition IDG trees in terms of CiM-supported instruction, and extract the groups that conforms to the offloading patterns;
\label{selection}
\end{algorithm}


To support IDG construction and data locality identification, we collect a set of data as given in Table~\ref{inststate} for all the instructions in the committed instruction queue (CIQ) as only committed instructions are important for program execution. We refer to these data as {\em instruction state (I-state)} which can be collected from both CPU and memory as shown in Fig. \ref{fig_simulation} (more details in Section \ref{sec_modeling}). The first three terms in I-state describe when and where an instruction is committed and executed, while the last tree terms detail the memory level as well as its execution status for a memory access/request instruction. Algorithm \ref{selection} summarizes the process of selecting offloading candidates when the I-state information is ready. Details about the construction of the various tables and the IDG are given in the next subsection.

\begin{table}[tb]
    \caption{I-state specification.}
    \label{inststate}
    \begin{center}
        \begin{tabular}{|c|p{5cm}|}
            \hline
            I-state element & Definition \\ \hline
             \multirow{2}{*}{Sequence index} &  Location of the instruction in the committed instruction queue CIQ \\ \hline
             Mnemonic code & Assembly code for each instruction  \\ \hline
             \multirow{1}{*}{Execution logic} &  Triggered functional unit that executes the instruction\\ \hline
             \multirow{2}{*}{Request from master} & Request address range of a load instruction and its issuing time
            \\ \hline
             \multirow{2}{*}{Memory access} & Address range of accessed memory objects (cache and main memory)
            \\ \hline
             \multirow{1}{*}{Response from slave} & Hit/miss status of each memory access
            \\ \hline
        \end{tabular}
    \end{center}
\end{table}


\begin{figure*}[htb]
    \centering
    \includegraphics[width=18cm]{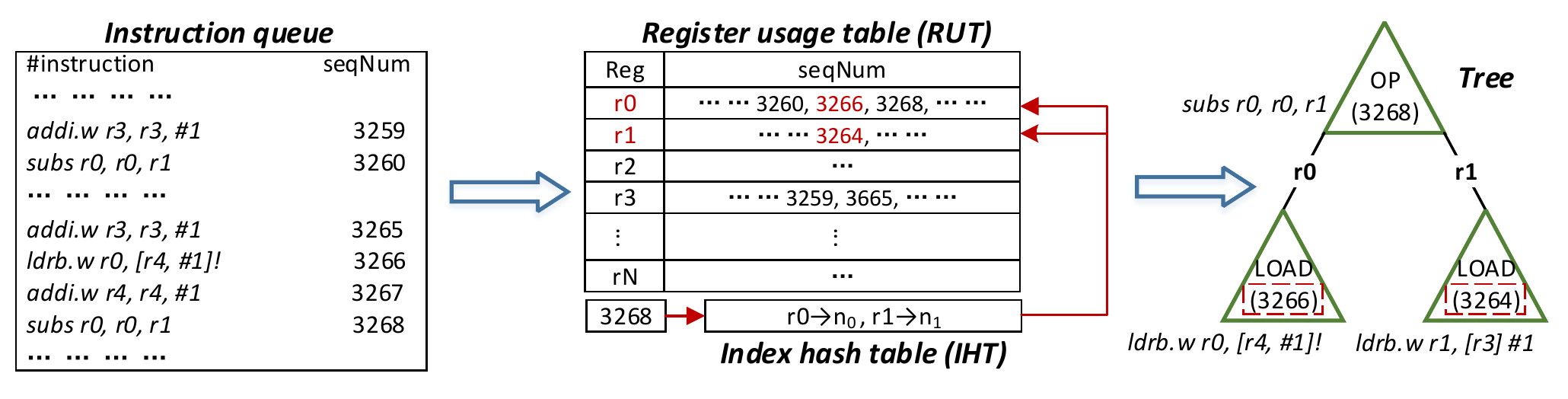}
    \caption{Procedure for IDG tree construction: (a) Instruction queue; (b) RUT and IHT; (3) IDG tree.}
    \label{dependency}
\end{figure*}

\eat{\subsubsection{Information Pre-processing}
\label{subsec_info_proc}
For a given application, the modeling flow in Section \ref{sec_modeling} collects the probed information of instruction sequences, pipeline execution, memory request and response. For probed instruction and pipeline traces, we can easily extract operator, operands and sequence index of an instruction from the instruction queue, which is sorted per its commit order. Meanwhile, the extraction of data information involves hit and miss from the memory hierarchy and happens to be more complex. As shown in Fig. \ref{memorytrace}, each request or access is associated with a specific address range for the requested or accessed data. For the request packet sent by the LSQ, we need to check if the request address is within the access address and then obtain the corresponding MSHR state. We can do such a procedure repeatedly until we find out the level in the hierarchy that stores the data.}


\begin{algorithm}[htb]
\caption{Algorithm for IDG tree construction.}
\setlength{\parindent}{0pt}
	$\textbf{Procedure:}$ IDG\_tree\_construction \\
	$\textbf{Input:}$\emph{instruction queue $Q$, CiM-supported instruction set $CiMSet$}, $RUT$, $IHT$\\
	$\textbf{Output:}$ IDG tree \emph{Tree}\\
	1. {\textbf {for}}  $k_{th}$ instruction $I_k$ in $Q$ {\textbf {do}}\\
	2. \quad {\textbf {if}} operation type of $I_k$ in $CiMSet$\\
	3. \qquad initialize a $subTree$ with $I_k$ as the root node\\
	4. \qquad $create\_tree(subTree)$\\
	5. \qquad append $subTree$ to $Tree$\\
	6. \quad \textbf{endif}\\
	7. \textbf{endfor}\\
	8. \textbf{return} $Tree$\\
	9. $\textbf{SubProcedure:}$ $create\_tree$($root$)\\
	10.\quad \textbf {if} $root.leftNode$==$NULL$ and $root$ is not a leaf node\\
	11.\qquad $j\leftarrow$ lookup $IHT$ by $root->seqNum$\\
	12.\qquad $root.leftNode\leftarrow$ lookup $RUT$ with [$j$]\\
	13.\qquad \textbf{if} operation type of $leftNode$ is Load\\
	14.\qquad \quad $root.leftnode=LEAF\_TRUE$\\
	15.\qquad \textbf{endif}\\
	16. \quad \textbf{endif}\\
    17.\quad \textbf {if} $root.rightNode$==$NULL$ and $root$ is not a leaf node\\
	18.\qquad $j\leftarrow$ lookup $IHT$ by $root->seqNum$\\
	19.\qquad $root.rightNode\leftarrow$ lookup $RUT$ with [$j$]\\
	20.\qquad \textbf{if} operation type of $rightNode$ is Load\\
	21.\qquad \quad $root.rightnode=LEAF_TRUE$\\
	22.\qquad \textbf{endif}\\
	23. \quad \textbf{endif}\\
	24.\quad \textbf{if} $root.leftNode$\\
	25.\qquad $create\_tree(root.leftNode)$\\
	26. \quad \textbf{endif}\\
	27.\quad \textbf{if} $root.rightNode$\\
	28.\qquad $create\_tree(root.rightNode)$\\
	29.\quad \textbf{endif} 
\label{recursion}
\end{algorithm}

\subsection{IDG Construction}
\label{subsec_idg}

Here we present a method to reduce the effort and complexity of constructing the IDG for a given program.
It is noted that, with ``store" nodes in Fig. \ref{newpattern} removed, IDG simply consists of many flipped trees. Thus, we introduce a compact tree structure with the following restrictions to reduce the redundancy in IDG:
\begin{itemize}
    \item With ``store" node removed, ``OP'' instruction is the root of the tree and must be the operation that CiM supports.
    \item The left and right children of a node in the tree represent the instructions that feed source data to the node.
    \item The leaf node needs to be either a load instruction or an immediate value. 
    \item An offloading candidate can include one or more connected nodes in the same tree.
    \item The data of an offloading candidate need to be in the same memory bank.
\end{itemize}
Fig. \ref{dependency} demonstrates the procedure for tree construction. The instruction queue on the left of Fig. \ref{dependency} lists the instructions as well as their indices in the CIQ. 
In order to avoid the complexity of recursive search for IDG tree construction, we here introduce the concept of Register Usage Table (RUT), as shown in the middle of Fig. \ref{dependency}. RUT keeps track of the committed time ($i.e.$, sequence index defined in Table~\ref{inststate}) when a register is used as the destination operand. This is due to the fact that the two connected nodes in an IDG tree must share at least one register. Each row in RUT corresponds to one register and maintains a list of sequence indices of the instructions that use the register. Another auxiliary index hash table (IHT) is also used to keep track of the source operand information for an instruction, with each entry corresponding to an instruction in CIQ. IHT records the registers ($r_i$) used as source operands for an instruction and the corresponding location ($n_i$) of the register when the instruction information is added to RUT. 
When a CiM-supported instruction is added as a node to an IDG tree, we can use its sequence index and IHT to find its source registers. Then with RUT we can locate the instructions that commit the last use of those registers as destination, which are also the child nodes to be added to the tree. Algorithm \ref{recursion} summarizes the complete algorithm for tree construction. 
By repeating this procedure, we can then build the trees for IDG with a $O(N)$ complexity, where $N$ is the number of nodes in the trees. As on the right of Fig.~\ref{dependency}, each node in the tree contains the information of operator, operands, and its sequence index. 

For the example in Fig. \ref{dependency}, when the instruction indexed at 3268 is added to the tree, we can first find out its source registers of $r_0$ and $r_1$ through IHT, which also tells us the location that appears of $r_0$ and $r_1$ in RUT when the instruction is committed, $e.g.,$ $n_0$, $n_1$, respectively. Then in RUT, the $n_{0\textrm{th}}$ entry in the list for $r_0$ is just the last instruction that uses $r_0$ as the destination. In other words, the instruction indexed at 3266 is just the left child node to be added in the tree. The same procedure is repeated for the right child. Since the two nodes happen to be ``LOAD" operations, the tree terminates at those two leaf nodes, as shown on the right of Fig. \ref{dependency}. 

Then, for any application of interest, \pecim can track the virtual address with the proposed IDG, thereby identifying data locality for CiM. Other than that, the framework may also adopt the prior circuit and architecture level efforts \cite{hshieh16, aga17} that use address translation techniques or memory controller to satisfy the offloading condition, $i.e.$, data accessed by CiM instructions are located on same bit-line. For example, \cite{hshieh16} uses a translation mechanism to allocate specific data structures into contiguous regions within the virtual memory space, then maps them to the physical memory space to ensure that the original data reside in the same cache array. \cite{aga17} further improves cache organization and address translation for operand locality by modifying the cache controller design to deal with offloading address constraints. Since \pecim aims to provide a system level evaluation framework to discover how much potential the technology and architecture may benefit from CiM, we will not place our focus on circuit or architectural innovations, which can be enabled in \pecim with the corresponding architectural models to provide more detailed simulation and higher evaluation accuracy.

\subsection{Trace Reshaping for System Profiling}\label{subsubsec_trace}

After offloading candidates are determined, the last task of the analysis stage is to reshape the instruction trace to meet the demands of the profiling stage (to be discussed in detail in Section \ref{subsubsec_system}). The instruction trace reflects the actual execution flow of a program. First, we need to reallocate the execution of selected instructions to the corresponding level of memory where the source data reside. Second, we need to remove those selected offloading instructions from the pipeline, re-organize data locality in the memory and replace them with the corresponding CiM-instructions. The reshaped trace then contains both regular and CiM-supported operations. Through reshaping the instruction trace, all instructions are explicitly allocated to either the function units on the CPU or the CiM module at different levels in the memory hierarchy. Then the profiler (to be discussed in the next section) tracks the activities of both the CPU and the CiM module (including, $e.g.$, instruction types, ALU accesses, L1 hits/misses) to estimate the energy of each module as well as the overall system.

The remaining issue for reshaping is managing data locality and dependency. Note that only when all the operands are available in the same cache level, we can issue the operation to the cache sub-array. Otherwise, we need to write the operand at the higher-level cache back to the lower-level cache, and forward its operator to the same level \cite{aga17}. Fig. \ref{treeexample}(c) shows an example of data dependency that the output of one tree is the input to another. 
In \pecim, with a regular compiler, we introduce a post-processing step to approximately mimic the CiM behavior. \pecim first traverses all the trees in the post order to ensure the right execution sequence. Then if two sub-trees are exacted from the same IDG tree, \pecim combines them to one in-cache operation to move data and manages data locality within the bank.

%% file: modeling.tex
\section{Modeling and Profiling}\label{sec_modeling}
In this section, we present the details in the modeling and profiling stage. The modeling stage provides the instruction execution and memory access information for a given program. It also provides the CiM model data 
to the profiling stage. The profiling stage then uses the output from the analysis stage as well as the CiM module data to obtain the overall system energy consumption.

\begin{table}[t]
    \caption{Probes attached to CPU and memory.}
    \label{probe}
    \begin{center}
        \begin{tabular}{|c|p{5cm}|}
            \hline
             Probe name & Monitored object \\ \hline
            \multirow{2}{*}{InstProbe} &  Time and execution in terms of pipeline status for each instruction \\ \hline
            \multirow{2}{*} {PipeProbe} &  Statistics of triggered function units for completing one instruction in CPU  \\ \hline
            \multirow{2}{*} {RequestProbe} & Track of request packet transmitted from LSQ including its issue time and address \\ \hline
            \multirow{2}{*}{ AccessProbe} & Record of memory access including time, access object, and hit/miss status
            \\ \hline 
        \end{tabular}
    \end{center}
\end{table}

\subsection{Application Modeling} \label{subsec_app}

\begin{figure}[tb]
    \centering
    \includegraphics[width=8cm]{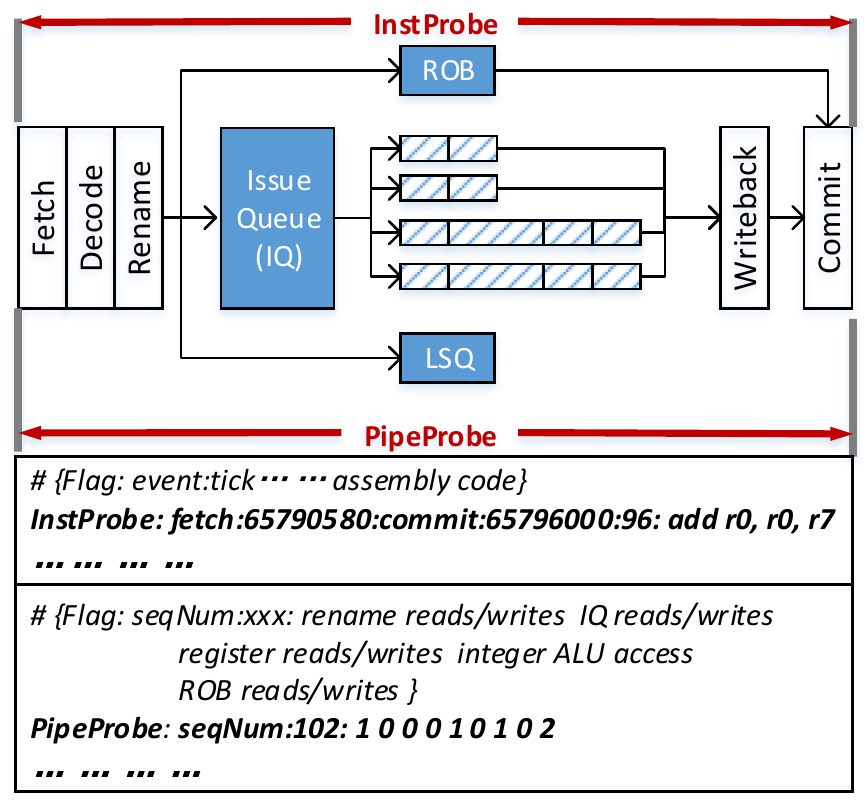}
    \caption{InstProbe and PipeProbe attached to an Out-of-Order CPU model.}
    \label{pipeline}
\end{figure}
As we stated earlier, application modeling aims to extract information about when and where instructions are executed and memory accesses occur. To be more precise, application modeling produces the I-state information (see Table \ref{inststate}) needed by the analysis stage.
we propose to leverage GEM5 augmented with carefully placed probes to obtain the I-state information. 

Specifically, Table \ref{probe} summarizes four probes as well as the monitored objects. InstProbe and PipeProbe monitor the execution status and triggered functions in the CPU, 
while RequestProbe and AccessProbe monitor memory behaviors. 
Below we discuss these two sets of probes in more details.

InstProbe collects time and execution in terms of pipeline status for each instruction, and PipeProbe collects which and when functional units are triggered by each instruction. There are two complications when collecting these data. First, when there are available resources for execution, multiple instructions are issued from Issue Queue (IQ) to several function units. Second, because of branch mis-prediction, only committed instructions are included in CIQ that is used for our offloading candidate analysis. Thus, these probes must be carefully placed to ensure correct information is collected.

To illustrate how these probes can be placed, we use the example of ARM ISA for a physical-register-file architecture with an out-of-order pipeline. Seven pipeline stages are executed in this architecture as shown in Fig. \ref{pipeline}. For each committed instruction, the InstProbe records the tick numbers of different pipeline stages according to the Programming Counter (PC) value. Meanwhile, the PipeProbe keeps tracks of the instruction  index in CIQ as well as the statistics of the triggered functional units ($e.g.$, IQ reads/writes, ROB reads/writes). The collected information by the two probes is processed for extracting the sequence index, assembly code and execution logic included in I-state. Then we utilize I-state to obtain the lifetime of an instruction and evaluate the overhead when an instruction is moved from CPU to CiM module.



\begin{figure}[tb]
    \centering
    \includegraphics[width=8cm]{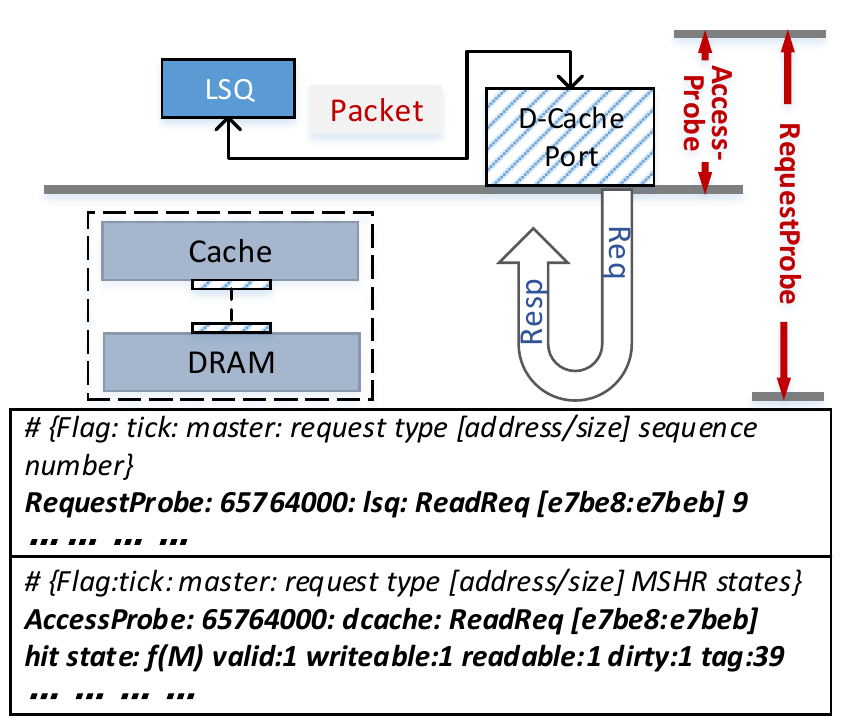}
    \caption{RequestProbe and AccessProbe for request packets monitoring.}
    \label{memorytrace}
\end{figure}

For RequestProbe and AccessProbe, Fig. \ref{memorytrace} describes where they are inserted as well as the information they collect. It is noted that the range of accessible addresses varies with the memory hierarchy level. Thus, a RequestProbe not only probes the tick of instruction execution, its master port, but also the address range of the ``Load" instruction. Similarly, an AccessProbe collects tick information, master port, hit or miss statistics of an address range, and status for Miss-status Handling Register (MSHR). 

The two probes can effectively capture the packets between the LSQ units and memory objects, so we can accurately obtain the access instruction and its request address. Once the packet is transported to the memory, we can track the packets among different levels in the memory hierarchy with response statistics and cache protocol. Apparently, the probed information depends on the application and the architecture, but is independent of the memory technology. 
\begin{table*}[tbp]
\newcommand{\tabincell}[2]{\begin{tabular}{@{}#1@{}}#2\end{tabular}}
    \caption{Cache energy (pJ) per operation in different configurations for SRAM and FeFET-based CiM architectures.}
    \label{cacheenergy}
    \begin{center}
        \begin{tabular*}{18cm}{@{\extracolsep{\fill}}lcccccccc}
            \hline\hline
            & Technology & Level & Config & \tabincell{c}{Non-CiM read} &  \tabincell{c}{CiM-OR} & \tabincell{c}{CiM-AND} & \tabincell{c}{CiM-XOR} & \tabincell{c}{CiM-ADDW32} \\ \hline
            & \multirow{2}{1em}{SRAM} & L1 & \tabincell{c}{4-way/64kB} & 61 & 71 & 72 & 79 & 79 \\
            &  & L2 & \tabincell{c}{8-way/256kB} & 314 & 341 & 344 & 365 & 365 \\ \hline
            & \multirow{2}{1em}{FeFET} & L1 & \tabincell{c}{4-way/64kB} & 34 & 35 & 88 & 105 & 105 \\
            &  & L2 & \tabincell{c}{8-way/256kB} & 70  & 72 & 146 & 205 & 205 \\ \hline
            \hline
        \end{tabular*}
    \end{center}
\end{table*}

\subsection{CiM Module Modeling} \label{subsec_device}

Besides the aforementioned application-related behaviors, the system-level benefits offered by CiM depend on the CiM construction and technology. 
A CiM module typically consists of a memory array and additional circuitry --- often present at the sense amplifier (SA) level --- responsible for generating output(s) that correspond to selective logical/ arithmetic operations.
SRAM-based caches that can perform bitwise AND, NOR, and XOR operations, among other computations, are proposed in \cite{aga17}. Alternatively, emerging NVMs have attractive features such as high density, low leakage power, low dynamic energy and fast access times, making them good candidates for the design of CiM main memories or caches. As pointed out in section \ref{sec_background}, STT-RAM, ReRAM, and FeFET-RAM are among the alternatives studied for the design of CiM architectures. Several CiM architectures proposed for NVMs also make use of a customized SA in a similar way to SRAM-based CiM approach \cite{jain18_sttcim,reis18_fefetcim,li16_pinatubo}. Among the aforementioned CiM architectures devised for NVMs, the FeFET-based is probably the most suitable for cache implementations due to its low write energy and latency as reported in \cite{reis18_fefetcim}. Thus, we pick SRAM- and FeFET-based CiMs as case studies for the proposed \pecim framework to be presented in section \ref{sec_design}. 

\begin{figure}[tb]
    \centering
    \includegraphics[width=5.5cm]{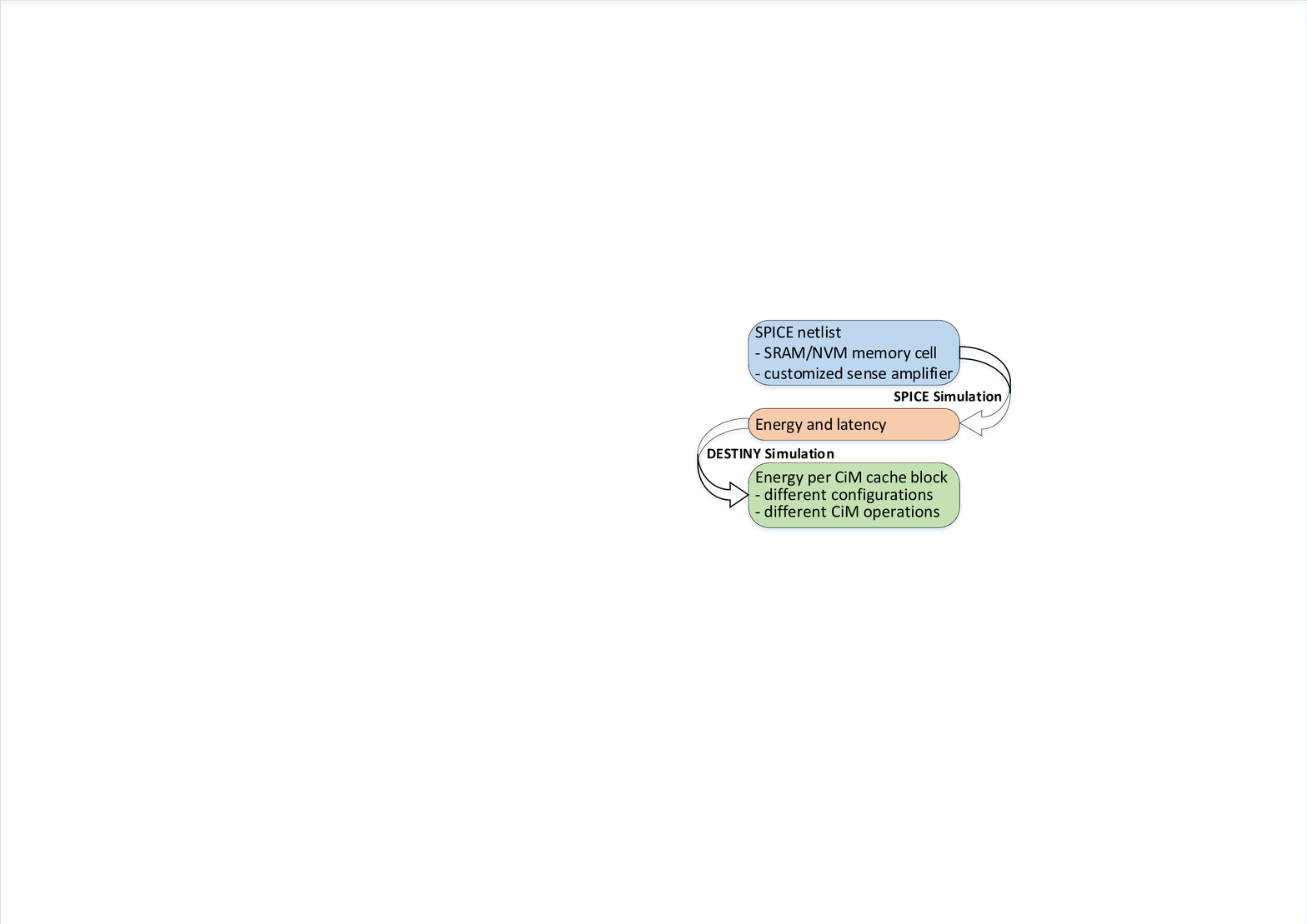}
    \caption{Flow for CiM module modeling for different operations.}
    \label{cachemodeling}
\end{figure}

Fig.~\ref{cachemodeling} illustrates our \textit{CiM module} evaluation flow. We employ the CMOS and FeFET SPICE models from \cite{vattikonda06,aziz16} to evaluate delay and energy of individual 6T-SRAM and 2T+1FeFET memory cells, as well as the customized SAs proposed in \cite{aga17, reis18_fefetcim}. To ensure a fair comparison between both designs, we (i) adopt the same technology node of 45nm in both designs, and (ii) port the full-adder part of the SA described in \cite{reis18_fefetcim} to the SRAM-based CiM \cite{aga17}. Thus, SRAM-based CiM and FeFET-based CiM can perform similar operations. We then employ SPICE-level results in a version of DESTINY \cite{destiny} that has been modified to support the evaluation of FeFET-based memories \cite{reis18_fefetcim}. 
DESTINY, as a microarchitecture-level tool for modeling 3D (and 2D) cache designs using SRAM, embedded DRAM, NVM ($i.e.$ STT-RAM, ReRAM, FeFET), has been widely validated against a large amount of multiple industrial prototypes in \cite{destiny}. Table \ref{cacheenergy} describes the energy data per operation ($e.g.$, non-CiM read, CiM read, AND, ADD, $etc.$) in different cache configurations obtained by the proposed models for both SRAM and FeFET-RAM, where non-CiM refers to regular operation in this paper. Although the focus of this work is placed upon SRAM and FeFET-RAM, other new technologies (and designs), such as RRAM, can be readily supported, as long as the latency and energy of each in-memory operation are specified.

%% file: profiling.tex
\textbf{}\subsection{Profiling} \label{subsubsec_system}
\subsubsection{Energy Evaluation}

\begin{figure}[tb]
    \centering
    \includegraphics[width=6.5cm]{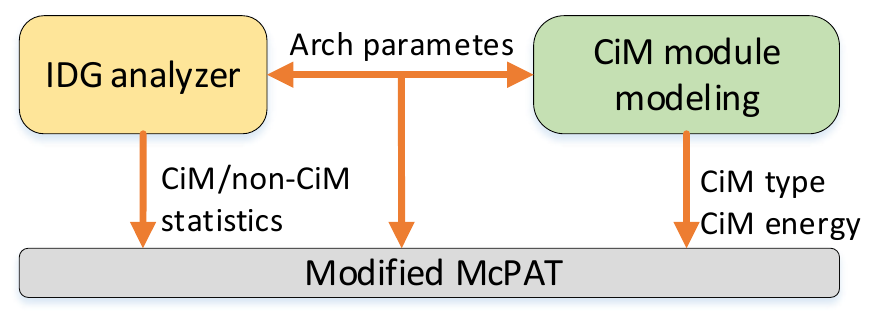}
    \caption{Architecture for CiM-enabled system profiler.}
    \label{profiler}
\end{figure}

Given the models and analyzer in the previous sections, we still need a system-wise profiler to combine the models at different design levels and report the overall system energy profile. Instead of building an energy model from scratch, we modify McPAT~\cite{mcpat} to evaluate the energy for both the CiM module and other functional units in a processor. Fig. \ref{profiler} shows the structure of our system-level profiler which relies on the application model (IDG), CiM model, architecture parameters, and modified McPAT. The original McPAT only computes energy and area for regular functional units using performance counter information (a set of statistics) extracted from an architectural simulator, or GEM5 in our work. In order to enable new CiM instructions, 
we employ the CiM model for CiM operations as discussed in the last subsection. Moreover, since some instructions are moved to the CiM module, we also need to reevaluate the energy of the host CPU. 

 We therefore modify and embed the following performance counters and models in McPAT: (i) Instruction type in pipeline and its count;
 (ii) Access time of function units in pipeline;
 (iii) Count of cache/DRAM read/write and hit/miss;
 (iiii) CiM operation type and its count.
Additional performance counters are added for CiM operations to ensure a unified energy model in the profiler. We can then safely invoke McPAT to use the modified performance counters and memory array parameters to estimate the energy consumption of the entire system. 

\subsubsection{Performance Evaluation}
\begin{figure}[tb]
    \centering 
    \includegraphics[width=8.5cm]{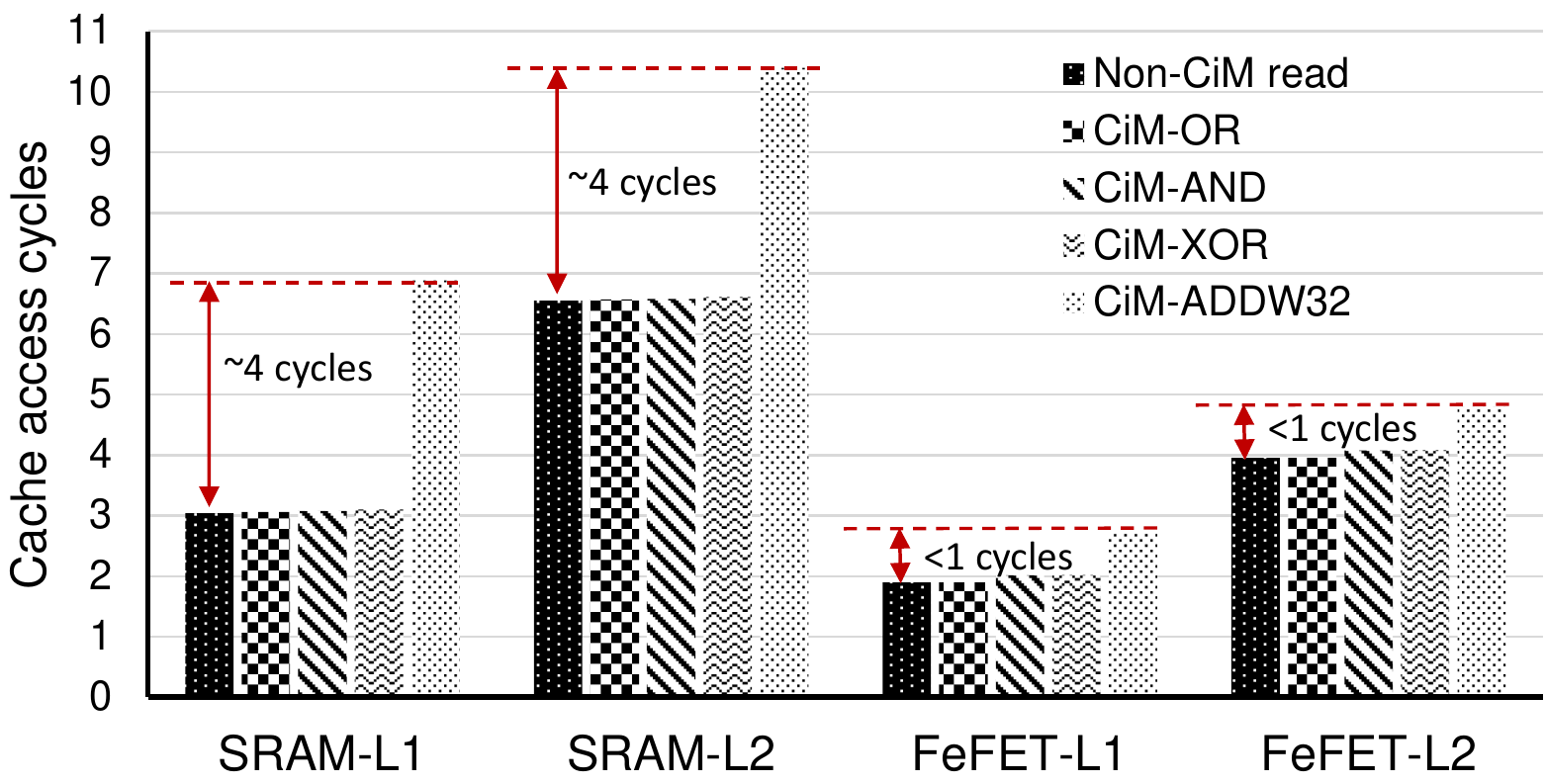}
    \caption{Access latency (cycles) of Non-CiM and CiM operations for SRAM and FeFET technologies.}
    \label{latency}
\end{figure}

Although CiM instructions migrate the workload of CPU, they incur additional memory access time. As reported in \cite{reis18_fefetcim}, CiM operations ($e.g.$, AND, ADD, $etc.$) may consume more time than regular data access due to the internal data migration and logic operations. \eat{At system level, the latency of a complete `load' instruction is largely impacted by whether the cache access is hit or not.} Thus, to estimate the system performance, we need to understand the impacts of instruction offloading on both the CiM module and host CPU. In particular, we need to extract the offloading ratio, cycle per instruction (CPI) and access latency of CiM operations to calculate the total time.  

\pecim records the execution profile of an instruction stream, which includes a fraction of instructions that have been committed on CPU but now are transferred to the CiM module. In a typical pipelined processor, there are two types of stalls that may impact CPI, memory access and pipeline bubbles. While the number of stalls due to memory access are changed when CiM instructions are used to replace regular instructions, the number of pipeline stalls may actually be reduced for CiM, with some instructions transferred to the CiM module. However, such difference is rather small and averaged out over the entire program trace. So we assume that while some instructions are skipped from CPU and added as CiM instructions, the system may keep constant CPI or execution efficiency.

Then the remaining parameters to be extracted are the access latencies for CiM operations. In \pecim, we employ HSPICE and DESTINY to estimate the latency of CiM and non-CiM operations. With the system clock frequency set at 1 GHz and the cache configuration the same as Table~\ref{cacheenergy}, Fig.~\ref{latency} shows the found access cycles of SRAM and FeFET for the particular operations and configurations. It is noted that the memory access time (or access cycles for a given clock frequency) in Fig.~\ref{latency} is the time that the operand is read or computed. For SRAM cache, the difference in access latency between non-CiM (the first operation in the figure, $i.e.,$ read) and CiM logic operations (second to forth operations in the figure, $i.e.$, OR) is almost negligible. Since the actual time to complete an access instruction includes data transfer time in the memory hierarchy, access time, and even re-access time if data hit fails, we may safely ignore the subtle difference between those operations and treat them with the same latency in  \pecim. On the other hand, CiM ADD operation in a cache bank takes almost four more cycles to complete than non-CiM read, which may result in severe pipeline stall. \pecim counts such actual access time for CiM ADD instruction when profiling the system.

\eat{Based on the CiM and non-CiM modeling data, we then can compare the latencies of non-CiM read and CiM operations conducted at different cache levels for a given technology. }

%% file: exploration.tex
\begin{table}[tb]
    \caption{Benchmark applications.}
    \label{table_benchmark}
    \begin{center}
        \begin{tabular}{|c|p{5cm}|}
            \hline
             Category & Application \\ \hline
             \multirow{3}{*}{Machine learning} & Naive bayes (NB), decision tree (DT), support vector machine (SVM), linear regression (LiR), Kmeans (KM)  \\ \hline
             \multirow{1}{*}{String processing} &  Longest common subsequence (LCS) \\ \hline
             \multirow{1}{*}{Multimedia app.} & MPEG-2 decode (M2D)\\ \hline
             \multirow{5}{*}{Graph processing} & Breadth first search (BFS), depth first search (DFS), betweenness centrality (BC), shortest path (SSSP), connected cononent (CCOMP), page rank (PRANK)
            \\ \hline
            \multirow{1}{*}{SPEC 2006} & Astar, H264ref, Hmmer, Mcf
            \\ \hline 
        \end{tabular}
    \end{center}
\end{table}

\section{{Experimental Results}} \label{sec_design}

We begin this section by first comparing the results obtained from \pecim with those obtained by DESTINY \cite{destiny} and \cite{jain18_sttcim} to help validate \pecim. We then present a number of simulation studies conducted with \pecim to demonstrate its capabilities and to provide insights on how various factors influence the benefits that a program can get from a CiM module. With \pecim , designers can explore the CiM based architectures with different design options, 
and then answer three key design questions raised in section I.\eat{ To present the effectiveness and potential of \pecim, this section also discusses the impacts of benchmarks to define CiM-favorable applications, the impact of system configurations ($i.e.$ cache size and associativity) and memory technologies ($i.e.$ CMOS, FeFET). } 
Note that our goal is not to highlight the benefits of CiM, which has already been shown in prior works. Instead, we aim to investigate the pros and cons from a system perspective regarding performance and energy consumption, thereby gaining insights in design tradeoffs for CiM based systems.

All the experiments are based on ARM Cortex A9, out-of-order core, 1.0 GHz system clock, with 512 MB main memory. The cache is configured with different capacities and associativities in our experiments. Here we use the CMOS SRAM as in \cite{aga17} for CiM implementation, in which all levels of the cache hierarchy are capable to conduct CiM operations. Our experiments employ 17 benchmarks from a wide range of application based on prior works \cite{jain18_sttcim, tesseract_isca_15, aga17, Liu_MICRO_2018, li16_pinatubo}, as summarized in Table \ref{table_benchmark}. They are considered as representatives of many typical accelerator workloads 
to stress \pecim~’s modeling and profiling capabilities across various dimensions.
\eat{
\begin{itemize}
    \item \textbf{Machine learning} includes a number of conventional machine learning algorithms.
    \item \textbf{String processing} chooses LCS, which is to find the longest subsequence common to all sequences in a set of sequences.
    \item \textbf{Multimedia application} includes the M2D benchmark which is highly iterative. 
    \item \textbf{Graph processing} employs commonly used graph processing algorithms for various applications.
    \item \textbf{SPEC 2006} selects several data intensive workloads from the classical CPU benchmark suite.
\end{itemize}}

\begin{figure}[tb]
    \centering
    \includegraphics[width=7.5cm]{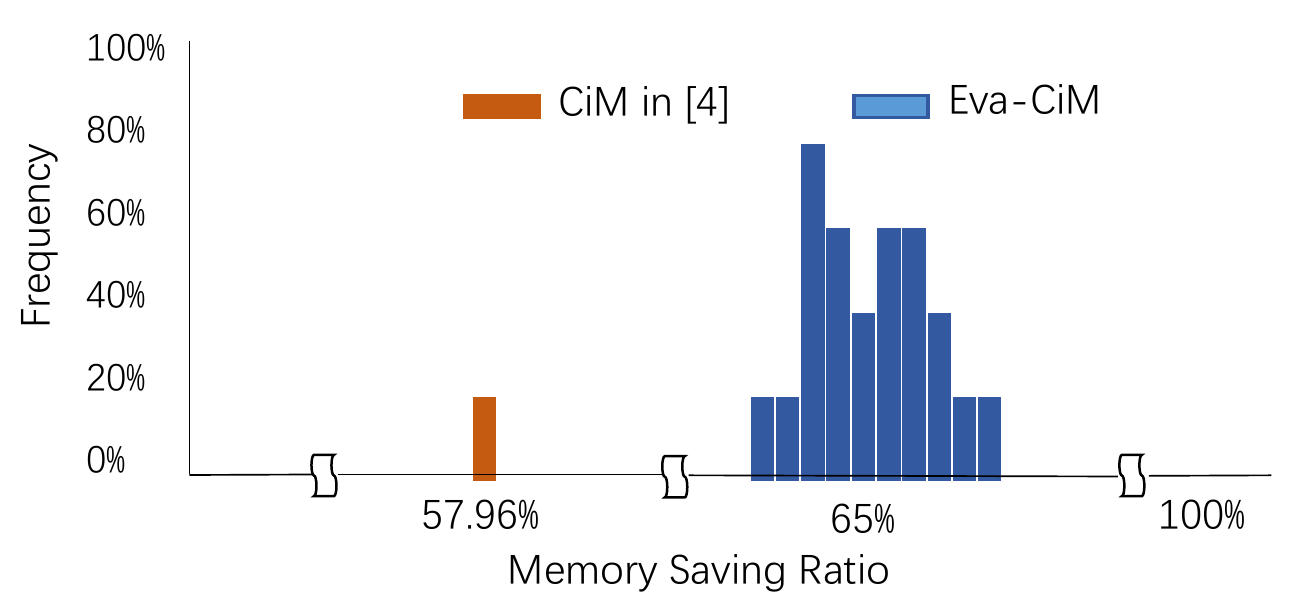}
    \caption{Comparisons on the CiM-supported memory accesses between \pecim and \cite{jain18_sttcim}.}
    \label{validation}
\end{figure}



\subsection{{Comparison and Validation}}

\begin{table}[t]
    \caption{Energy model comparison with DESTINY \cite{destiny}.}
    \label{table_energy}
    \begin{center}
        \begin{tabular}{|c|c|c|}
            \hline
             \multirow{2}{*}{Model} & \multicolumn{2}{c|}{Energy (nJ)}  \\ \cline{2-3}
             & CiM  & non-CiM  \\ \hline
             DESTINY \cite{destiny} & 455.49 & 124.43 \\ \hline
             \pecim & 565.18 & 154.40 \\ \hline
             Deviation & 24.0\% & 24.0\%\\ \hline
        \end{tabular}
    \end{center}
\end{table}

\begin{table*}[t]
\newcommand{\tabincell}[2]{\begin{tabular}{@{}#1@{}}#2\end{tabular}}
    \caption{Speedup, energy improvement and improvement breakdown (between processor and cache) for CiM-based $v.s.$ non-CiM systems.}
    \label{system_eval}
    \scriptsize
    \begin{center}
        \begin{tabular}{p{0.4cm}p{1cm}ccccccccccccccccc}
            \specialrule{0.05em}{0pt}{0.5pt}
            \specialrule{0.05em}{0.5pt}{5pt}
            \multicolumn{2}{c}{\textbf{\small{Benchmark}}} & NB & DT & SVM & LiR & KM & LCS & M2D & BFS & DFS & BC & SSSP & CCOMP & PR & astar & h264ref & hmmer & mcf  \\
             \specialrule{0.05em}{3pt}{3pt}
             \multicolumn{2}{c}{\textbf{\small{Speedup}}} & 1.51 &  1.52 & 1.42 & 1.24 & 1.30 & 1.31 & 1.34 & 1.40 & 1.55 & 0.99 & 1.34 & 1.52 & 1.42 & 1.28 & 1.17 & 1.36 & 1.27 \\ 
             \specialrule{0.1em}{3pt}{3pt}
             \multicolumn{2}{c}{\tabincell{c}{\textbf{\small{Energy}} \\ \textbf{\small{Improvement}}}} & 3.28 & 5.12 & 2.83 & 2.68 & 3.21 & 4.31 & 4.85 & 2.33 & 1.98 & 1.30 & 2.33 & 3.46 & 4.54 & 5.26 & 2.05 & 2.87 & 3.58 \\
             \specialrule{0.1em}{3pt}{3pt}
             \multirow{2}*{\textbf{\small{Ratio}}} & \textbf{\small{Processor}} & 1.01 & 0.92 & 0.92 & 1.16 & 0.91 & 0.91 & 1.01 & 0.98 & 1.53 & 0.90 & 1.12 & 1.01 & 0.91 & 0.97 & 0.86 & 0.93 & 0.93 \\ 
             \specialrule{0em}{1pt}{1pt}
             & \textbf{\small{Caches}} & -0.01 & 0.08 & 0.08 & -0.16 & 0.09 & 0.09 & -0.01 & 0.02 & -0.53 & 0.10 & -0.12 & -0.01 & 0.09 & 0.03 & 0.14 & 0.07 & 0.07 \\
             \specialrule{0.05em}{3pt}{0.5pt}
             \specialrule{0.05em}{0.5pt}{0pt}
        \end{tabular}
    \end{center}
\end{table*}

\begin{figure}[tb]
    \centering 
    \includegraphics[width=8.5cm]{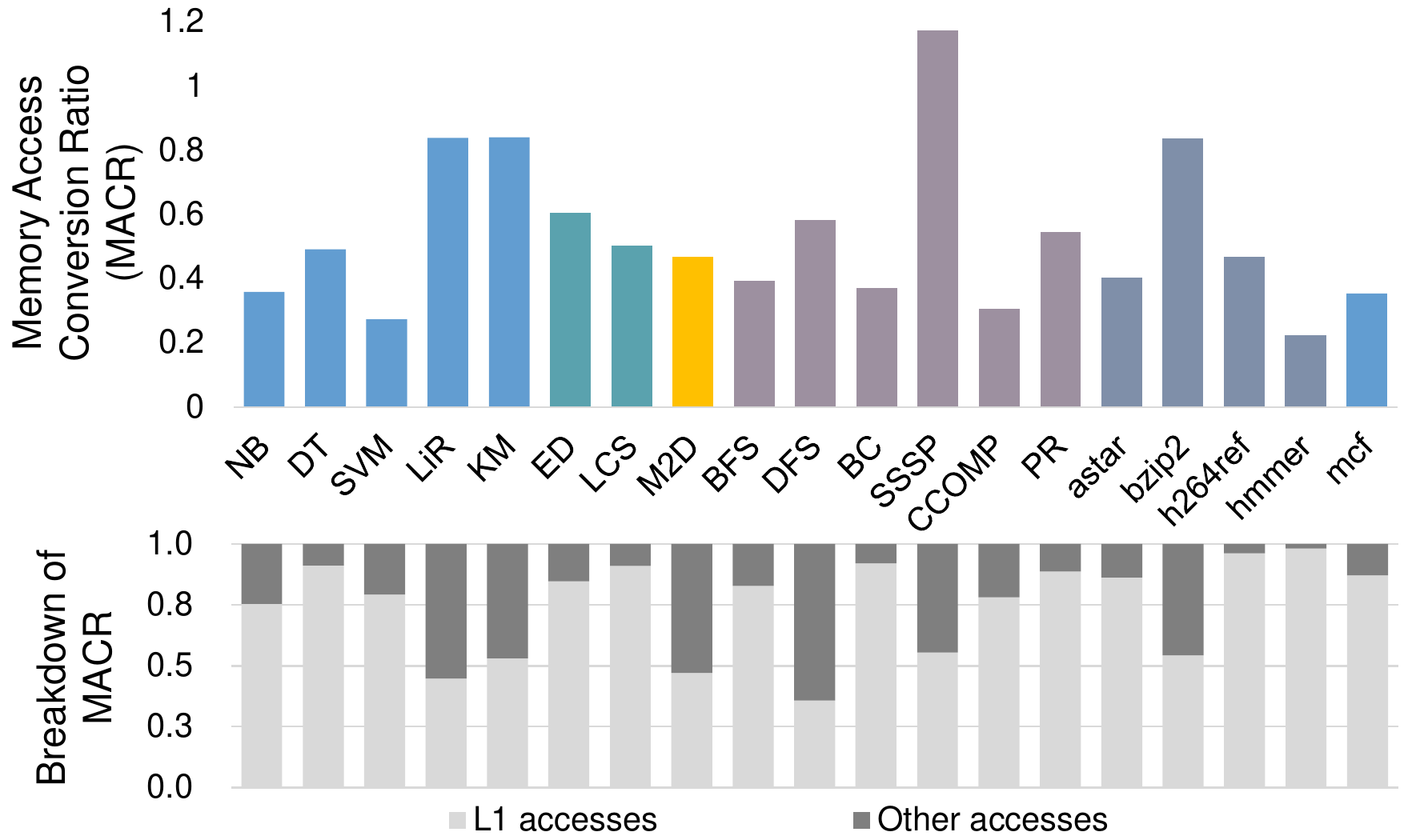}
    \caption{MACR for different benchmarking programs (top); Breakdown of MACR into L1 accesses and other accesses for different benchmarking programs (bottom).}
    \label{access_breakdown}
\end{figure}

The behavior of CiM depends on not only the benchmarks, but also the entire procedure of compiling, decoding and execution. Thus, the deviations in compiler, core architecture, or memory hierarchy may all impact the results. It is noted that much of the existing literature on CiM accelerators only focuses on the design and optimization of the computational components and internal memory for the CiM module while assuming that all the data needed already resides in CiM. Instead, \pecim is designed  to assess how data movement and interactions affect CiM design system.

We choose to conduct the validation by comparing the two major parts of \pecim, energy estimation with DESTINY, and CiM operation count with \cite{jain18_sttcim}, using one application program, LCS. We adopt the same experiment setup for all the tools for fair comparison, in which all the levels (consisting of 32KB/4-way L1 and 256KB/8-way L2) of cache hierarchy are capable to conduct CiM operations. 
 
We first use \pecim to obtain the energy consumption of a trace of LCS with around 3000 instructions. Then we use DESTINY and a modified McPAT to estimate the energy of these instructions as in Sec.~\ref{subsec_device} and \ref{subsubsec_system}.  As showned in Table~\ref{table_energy}, the energy estimates by the two approaches show around 24\% difference for both the CiM and non-CiM instructions. This difference is reasonable since, though \pecim employs DESTINY for per-operation energy estimation, it also accounts for the impact of multi-level cache hierarchy, such as cache access miss, $etc.$. 
For performance comparison, since \cite{jain18_sttcim} uses an emulation platform with a simplified in-order processor as well as 1 MB SPM, we modify the evaluation architecture accordingly with a cache size of 1MB. 
Note that the comparison focus is placed upon the count of instructions that are offloaded to the CiM module. 
We execute the LCS code 20 times with randomly generated inputs and breakdown  memory accesses using a similar approach as in \cite{jain18_sttcim}. As illustrated in the histogram on the right of Fig.~\ref{validation}, \pecim selects around 65\% memory accesses for offloading to CiM while \cite{jain18_sttcim} reports 58\%. This discrepancy is mainly due to the differences from the two underlying ISAs and higher complexity of the memory hierarchy used in \pecim than the SPM structure used in \cite{jain18_sttcim}.

In summary, both energy and performance comparisons with existing works show that \pecim results are close enough to gives us confidence in the effectiveness of \pecim.
We will further use \pecim to investigate the impact of CiM module on energy and performance as well as various design options.

\begin{figure*}[htb]
    \centering 
    \includegraphics[width=16cm]{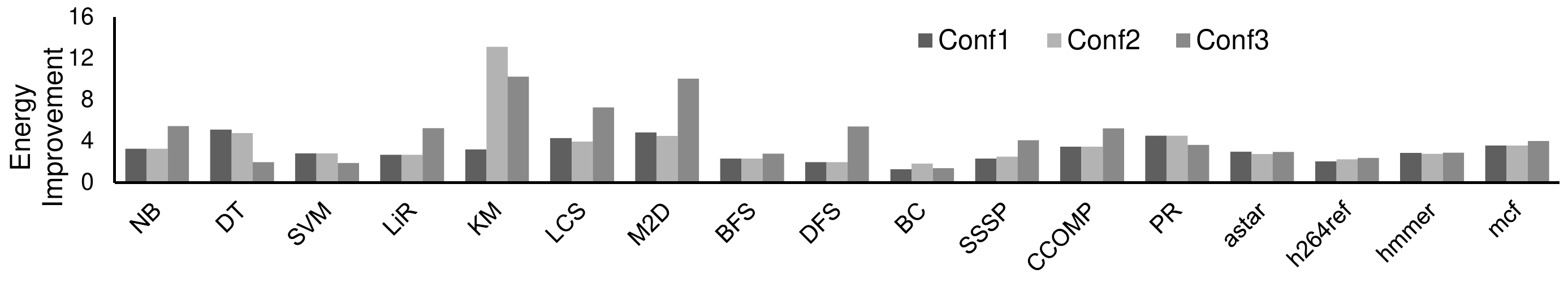}
    \caption{Energy improvements for CiM with different cache configurations. }
    \label{conf}
\end{figure*}

\begin{figure*}[htb]
    \centering 
    \includegraphics[width=16cm]{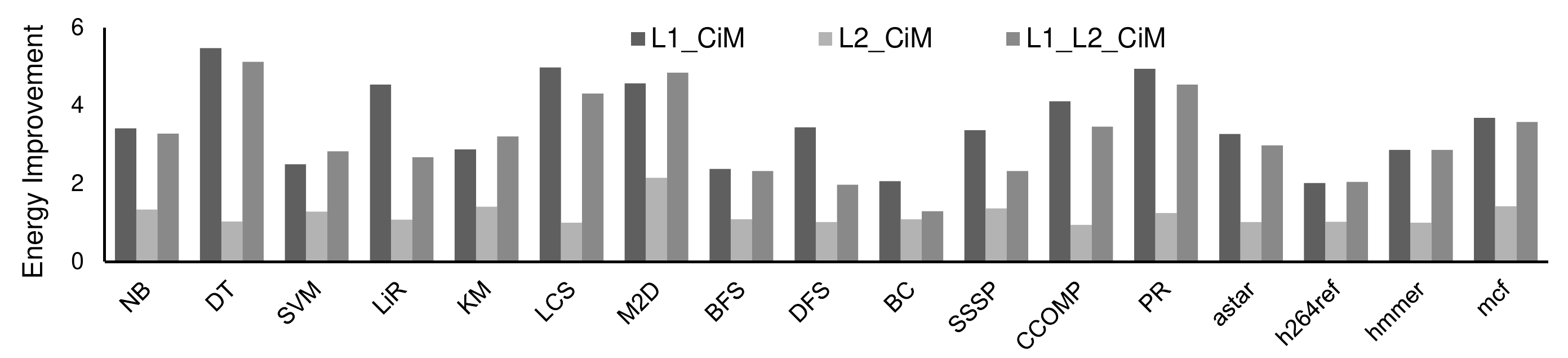}
    \caption{Energy improvement comparison among CiM supported by only L1, CiM supported by only L2 and CiM supported by both L1 and L2.}
    \label{CiMconf}
\end{figure*}

\subsection{{Performance and Energy Evaluation for System with and without CiM Module}}\label{eval_approach}
\pecim provides a flexible, modular simulation framework that makes it possible to explore CiM architecture by offering a diverse of evaluation capabilities. This subsection aims to show that the tool can provide in-depth investigations to evaluate whether it is beneficial to include a CiM module in the system, which is a key question many designers are interested in. 
 
We start our investigations with the performance comparison. The speedup of CiM over the non-CiM system is shown in the second row of Table~\ref{system_eval}, which ranges from 1.0-1.5$\times$ for various benchmarks. Meanwhile, it is noted that the performance of some benchmarks actually degrades, $e.g.$, BC in Table~\ref{system_eval}.

We then evaluate the total energy including both host CPU and cache for the aforementioned application benchmarks and then report the energy improvements. We focus on energy effect of cache caused by CiM without considering subtle influence on the host CPU, then compute the ratio between energy variation and baseline energy.
The third row of Table \ref{system_eval} shows the total \textit{energy improvements}, defined as \textit{the ratio of the baseline system energy (without a CiM module) over the one when with a CiM module}. The energy improvements of the benchmarks range from 1.3-6.0$\times$ for various applications. Energy improvements are contributed by both the CiM module (cache) and host CPU, and their breakdowns are shown in the last two rows of Table~\ref{system_eval}. Note that the energy improvement is mainly contributed by the host side, which is expected due to the reduced number of memory accesses. 
As a result, across the set of benchmarks, we have mixed results: some programs show positive energy improvement contributions from the CiM module while some have negative contributions.

To help understand the experimental results, we first introduce two terms: \textit{CiM-favorable} and \textit{CiM-unfavorable}. CiM-favorable programs tend to achieve greater energy improvement from CiM. For example, DT, LCS, PR, and astar, can be classified as CiM-favorable programs and hence more suitable for CiM-based systems. On the other hand, CiM-unfavorable programs ($e.g.$, LiR, DFS) receive much limited improvements.

\subsection{{Impact of Benchmarks}}
\pecim can be used to study the program's characteristics that influence whether a program is CiM-favorable or not. In many prior works with non-cache-able memory, $e.g.$, \cite{zhang14, li16_pinatubo}, any problems with significant memory accesses that have good data locality are considered to be CiM-favorable. 
However, very few provide detailed breakdown of memory accesses, especially when the CiM module functions as a general purpose computing block. Due to the system complexity, multi-level memory hierarchy and lack of CiM-centric compiler support, it is possible that not all data locality can be exploited by CiM as assumed by prior work. We have conducted experiments on the given application programs to investigate the percentage of instructions that have the ``proper" data locality to allow the associated operations to be migrated to the CiM module. 

To capture this concept of ``proper'' data locality, we introduce a metric called \textit{memory access conversion ratio} (MACR), which is the ratio between the accesses with appropriate locality that can be replaced by CiM operations and the regular memory accesses. Fig.~\ref{access_breakdown} presents the breakdown of memory accesses according to MACR. The results clearly show that, for a given system architecture with a specific CiM design, MACR can be smaller than one even for the programs that are commonly considered as data-intensive, $e.g.$, M2D. For such cases, CiM actually provides relatively low energy improvements, as shown in Table~\ref{system_eval}. Based on the data shown in Fig.~\ref{access_breakdown} and Table~\ref{system_eval}, it can be seen that a high MACR ($e.g.$, 50\% or more) is an indicator for a program to be CiM-favorable.

\subsection{Impact of System Configuration and Architecture}
\pecim helps designers study the impact of system configuration and architecture on a CiM system. 
Fig. \ref{conf} illustrates the results for different cache configurations. Here we have three configurations: (i) 32KB/4-way L1 and 256KB/8-way L2, (ii) 64KB/4-way L1 and 256KB/8-way L2, (iii) 64KB/4-way L1 and 2MB/8-way L2. It is clear that most applications ($e.g.$, NB, LCS, SSP, $etc.$) experience higher benefits for larger cache sizes. However, it is also noted that while a larger cache size helps CiM, the energy per operation is also increased (as shown in Table \ref{cacheenergy}), which actually reduces the benefit from CiM. 

In addition, we investigate the impact of different cache levels that support CiM. Fig. \ref{CiMconf} depicts the results of energy improvements when CiM instructions are supported by L1 only, L2 only, and both of them. We use 32KB/4-way L1 and 256KB/8-way L2 for CiM implementation. In general, applications exhibit lower energy improvements when CiM is only supported by L2, which is due to the more frequent L1 access in a system with complete memory hierarchy as well as smaller energy overhead for CiM operations in L1. 

\begin{figure*}[htb]
    \centering 
    \includegraphics[width=16cm]{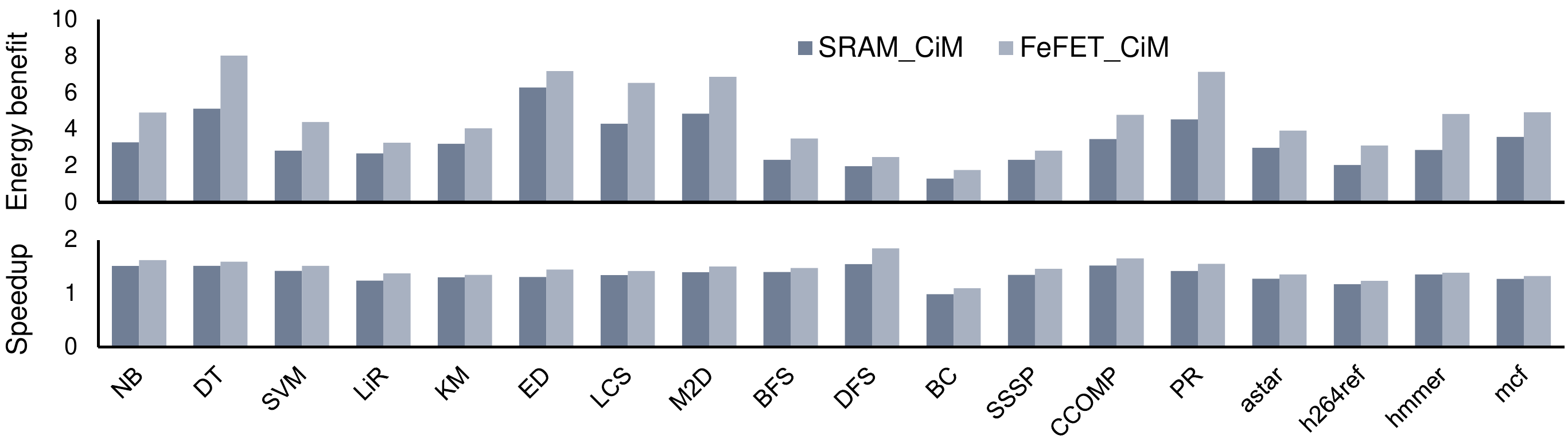}
    \caption{Benefits for CMOS SRAM v.s. FeFET RAM: Energy improvement (top); Performance improvement (bottom).}
    \label{FeFET}
\end{figure*}

\subsection{{Impact of Technology}}
We utilize \pecim to explore the performance and energy benefits when using different device technologies for CiM. Due to the flexibility of the profiler, \pecim supports multiple memory technologies given the device parameters that can be provided by circuit modeling as in section~\ref{subsubsec_system}. Here we present the performance and energy comparison between CMOS SRAM and FeFET-RAM in Fig.~\ref{FeFET}. The energy improvements are normalized to the non-CiM baseline system using CMOS SRAM. We observe that the energy benefits for FeFET based CiM is about 50-70\% higher, and are consistent across all the benchmarks. Additionally, FeFET SRAM outperforms CMOS SRAM in terms of performance due to its lower latency of CiM operations. Thus, \pecim may provide researchers with the capability of design space exploration to make better trade off among different technologies.